\documentclass[12pt]{iopart}
\usepackage[normalem]{ulem}  
\usepackage{graphicx}
\usepackage{capt-of}
\usepackage{amssymb}
\usepackage{mathabx} 
\usepackage{lineno}
\usepackage{multirow,tabularx}
\usepackage{utfsym}

\graphicspath{{img/}}

\begin{document}

\title[APOLLO: millimeter accuracy dataset characterization]{Fifteen years of millimeter accuracy lunar laser ranging with APOLLO: dataset characterization}

\author{J.B.R.~Battat$^1$, E.~Adelberger$^2$, N.R.~Colmenares$^{3,4}$, M.~Farrah$^1$, D.P.~Gonzales$^3$, C.D.~Hoyle$^5$, R.J.~McMillan$^6$, T.W.~Murphy,~Jr.$^3$, S.~Sabhlok$^3$, C.W.~Stubbs$^7$}
\address{$^1$ Department of Physics, Wellesley College, 106 Central St, Wellesley, MA 02481, USA}
\address{$^2$ Center for Experimental Nuclear Physics and Astrophysics, Box 354290, University of Washington, Seattle, WA 98195-4290, USA}
\address{$^3$ Center for Astrophysics and Space Sciences, University of California, San Diego, 9500 Gilman Drive, La Jolla, CA 92093-0424, USA}
\address{$^4$ Geodesy and Geophysics Lab, NASA Goddard Space Flight Center, 8800 Greenbelt Rd., Greenbelt, MD, 20771}
\address{$^5$ Department of Physics and Astronomy, California State Polytechnic University, Humboldt, One Harpst St, Arcata, CA 95521-8299, USA}
\address{$^6$ Apache Point Observatory, 2001 Apache Point Rd, Sunspot, NM 88349-0059, USA}
\address{$^7$ Department of Physics, Harvard University, 17 Oxford St, Cambridge, MA 02318, USA}
\eads{\mailto{jbattat@wellesley.edu}}

\maketitle

\begin{abstract}
 We present data from the Apache Point Observatory Lunar Laser-ranging Operation (APOLLO) covering the 15-year span from April 2006 through the end of 2020. APOLLO measures the earth-moon separation by recording the round-trip travel time of photons from the Apache Point Observatory to five retro-reflector arrays on the moon. The APOLLO data set, combined with the 50-year archive of measurements from other lunar laser ranging (LLR) stations, can be used to probe fundamental physics such as gravity and Lorentz symmetry, as well as properties of the moon itself. We show that range measurements performed by APOLLO since 2006 have a median nightly accuracy of 1.7\,mm, which is significantly better than other LLR stations.
\end{abstract}

\noindent{\it Keywords\/}: lunar laser ranging, Moon, gravitation

\section{Introduction}

Lunar Laser Ranging (LLR) has provided high-precision measurements of the earth-moon separation since 1969 when Apollo 11 astronauts placed the first corner-cube retroreflectors on the lunar surface.
LLR measures the round-trip travel time of laser pulses from a telescope on Earth to a lunar reflector array. 
Five reflector arrays currently sit on the moon -- three deployed by Apollo astronauts and two on Soviet rovers.  Over the past half-century, LLR data precision has improved from a few hundred millimeters to a few millimeters.

The fifty-year-long span of LLR data provides precise determination of Earth orientation parameters \cite{muller2015}, the secular increase in the earth-moon distance \cite{2016LPI....47.1096W}, and the physical properties of the moon \cite{2015JGRE..120..689W}, including the discovery of a fluid lunar core \cite{2001JGR...10627933W}.  LLR data have also enabled some of the most precise tests of fundamental physics such as constraints on fifth-forces \cite{isl2003}, the time evolution of Newton's constant $G$ \cite{2010A&A...522L...5H}, strong equivalence principle violation \cite{Muller:2012sea,Williams:2012nc}, and Lorentz symmetry violation \cite{Battat:2007uh}. A recent review article provides a more complete overview of LLR's science reach \cite{2019JGeod..93.2195M}.
  
In 2006, a new LLR station called the Apache Point Observatory Lunar Laser-ranging Operation (APOLLO) began its measurement campaign at the Apache Point Observatory (APO) 3.5-meter telescope in New Mexico (USA) \cite{murphyAPOLLO2008, battatPASP2009}. Since then, APOLLO has provided millimeter-precision lunar range measurements \cite{apolloCQG2012,Liang:2017xjb}, which constitutes an order of magnitude improvement over the former state of the art, and a few parts in $10^{12}$ of the earth-moon distance. APOLLO members have contributed to the science of solar system tests of gravity \cite{Murphy:2007nt,2009SSRv..148..217M, Battat:2007uh,Battat:2008bu}, and other analysis groups have used the publicly available APOLLO data for gravitational physics constraints as well \cite{Muller:2012sea,Bourgoin:2016ynf}. 
  
APOLLO's exceptional instrumentation and associated return rate allowed lunar ranging studies that were not possible at other observatories. During most ranging sessions, APOLLO acquires signal from all five reflectors (Lunokhod 1 having been re-discovered by APOLLO in 2010 \cite{Murphy:2010ua}), which constrains the lunar orientation and tidal distortion that are needed to convert surface ranges into a measure of the lunar center-of-mass position -- itself critical for testing gravity. The high signal rate also leads to an accurate characterization of system throughput, from which we found evidence for degradation of reflector response, likely from lunar dust \cite{Murphy:2010az,Murphy:2013oya}.

  One of the main science drivers of LLR, and a motivating factor behind the creation of the APOLLO experiment in the first place, is the sensitivity of LLR to the strong equivalence principle (SEP) \cite{Nordtvedt:1968zz,Nordtvedt:1968qr,Nordtvedt:1968qs}. Until recently\footnote{An improved constraint on SEP violation was recently obtained from a hierarchical stellar triple system consisting of a white dwarf and millisecond pular orbiting another white dwarf \cite{Archibald:2018oxs}.}, in fact, the most stringent limits on SEP violation had come from LLR observations -- with limits on the differential acceleration $\Delta a$ of the earth and moon toward the sun at the level of $\left|\Delta a/a\right| = 10^{-13}$, which translates to a constraint on the SEP of $3\times 10^{-4}$ \cite{2019JGeod..93.2195M}. An EP violation would manifest in the LLR dataset as a perturbation to the earth-moon range at the 29.53-day synodic period of the moon \cite{Damour:1995gi,Damour:1996xt}. The phase of this signature gives a maximum range perturbation at new and full moon, with nulls at first and last quarter. Because of telescope Sun avoidance restrictions, ranging during new moon is not possible, and many stations have not been able to range at full moon either, for reasons described below. The LLR data set is thus sparse when the SEP signal is maximal. Prior to the start of APOLLO, it was recognized that more uniform coverage of the synodic month would improve the LLR-based SEP constraints, even absent any improvement in range precision \cite{1998CQGra..15.3363N}. The APOLLO experiment was designed for full-moon ranging and can do so (see Section~\ref{sec:lunarphase}). However, as described elsewhere \cite{Murphy:2013oya}, APOLLO's measured ranging throughput is an order of magnitude smaller during full moon than it is during the first and third quarter phases due to solar-radiation-induced thermal perturbations to the retro-reflector arrays.

While APOLLO and other LLR stations report their measurement precision based on photon statistics, their ranging \textit{accuracy} has traditionally been measured via comparison to a detailed Solar System ephemeris model. At present, all such models have known deficiences in excess of the millimeter-scale. Motivated by this, and by the desire to have an independent check on APOLLO's ranging accuracy, the APOLLO team built an Absolute Calibration System (ACS) in 2016 \cite{Adelberger:2017low}. The ACS enables direct, in-situ calibration of lunar range measurements and has since confirmed that APOLLO data accuracy is at the millimeter scale. Furthermore, we also have a mechanism to back-correct archival APOLLO data taken prior to the ACS at the $\sim 2$--$3$ millimeter level \cite{Liang:2017xjb}. Here, we report on the most recent APOLLO data release -- fifteen years of millimeter-accurate measurements of the earth-moon separation. As of January 2021, the operation of the APOLLO instrument has been transferred to NASA Goddard Space Flight Center, so this data release is the final one by the team that built the instrument.

In Section~\ref{sec:instrument}, we provide an overview of the APOLLO instrument and the ACS, including a list of important hardware changes made over the life of the instrument. Section~\ref{sec:data} provides statistics on the fifteen-year-long APOLLO dataset, situating it in the context of the rest of the half-century of LLR data from other stations. The main conclusions are that (i) APOLLO has produced 15 years of millimeter \textit{accurate} measurements of the earth-moon range, (ii) regular acquisition of three or more lunar retroreflector arrays during  a single ranging session improves constraints on lunar orientation, which in turn improves gravitational physics constraints, and (iii) APOLLO range measurements during full moon enhances LLR-based constraints on the equivalence principle. A detailed description of the APOLLO calibration and data reduction is presented in a companion paper~\cite{apolloReduction}.

\section{APOLLO and the Absolute Calibration System}\label{sec:instrument}
Full descriptions of the APOLLO apparatus and the ACS are provided elsewhere \cite{murphyAPOLLO2008,Adelberger:2017low}. Here, we give a brief overview of each at the level needed to interpret the dataset statistics that we report in Section~\ref{sec:data}. We also document all significant hardware changes made to the APOLLO system that could impact range measurement uncertainty.

APOLLO employs a frequency-doubled Nd:YAG laser (532\,nm wavelength) with a 20\,Hz repetition rate and 120\,ps pulse width (FWHM). The laser beam is coupled to the 3.5m telescope and emerges collimated from the primary mirror. A local corner cube (25\,mm diameter) attached to the telescope's secondary mirror intercepts a small portion of the outgoing beam and redirects it back to the receiver, a $4\times 4$ array of avalanche photodiode pixels each capable of sensing single photons. The arrival time of the light returning from the local corner cube constitutes the launch time for the lunar range measurement. Approximately 2.5\,s later, the detection of photons returning from the lunar reflectors produces receive times, giving lunar range measurements on a photon-by-photon basis. During off-line data analysis, individual range measurements taken over a time-span of $\sim 3$--8 minutes (3000--10000 laser shots) on a single lunar reflector array are reduced into a composite range measurement called a ``normal point'' (NP), which comprises a round-trip travel time at a representative launch time.
A companion paper provides a full description of the APOLLO data reduction pipeline and NP generation process~\cite{apolloReduction}.

The round-trip time is measured by a multi-scale timing system. Coarse timing is achieved by counting pulses from a 50\,MHz clock, while a 25\,ps bin-width time-to-digital converter (TDC) provides higher resolution timing. A stable frequency standard generates the 50\,MHz clock. Originally, APOLLO used the model XL-DC clock from Symmetricom as the frequency standard. The XL-DC maintains time in accordance with atomic standards to approximately 100\,ns by employing an ovenized quartz oscillator disciplined by reference to the global positioning system (GPS). 
In 2017, a cesium clock (5\,ps jitter) replaced the GPS-disciplined clock (7--10\,ps jitter) as the frequency standard for APOLLO. More importantly than low jitter, the cesium clock is more stable than the GPS-disciplined clock on 1000-second timescales where meteorology degrades the GPS clock accuracy.

The cesium clock also serves as the frequency standard for the ACS, which consists of a high-repetition-rate (80\,MHz), short-pulse-duration ($< 10$\,ps) fiber laser. These pulses are fed into the APD array along with the lunar launch and receive pulses, creating an overlaid ``ruler'' of optical ``tick marks'' 12.500\,ns apart. ACS measurements have confirmed that APOLLO ranges have an accuracy at the millimeter level, and a campaign that compared the GPS and cesium clocks provided a mechanism to correct for clock deviations and assign trustworthy range uncertainties to all APOLLO data since 2006 \cite{Liang:2017xjb}. 

APOLLO has undergone various hardware changes and upgrades over its lifetime, including changes to the system clock (see Table~\ref{tab:clock}). A comparison between the cesium clock (Cs, acquired for the ACS) and the XL-DC using a Universal Counter began in February 2016. This comparison enabled back-correction of the normal point uncertainties for data taken prior to installation of the ACS \cite{Liang:2017xjb}. In January 2017, we switched to using the Cs clock for the APOLLO time base. A problem with the Cs clock in February 2017 required that we revert to the XL-DC for a short time. In December 2017, the Cs clock experienced a second problem that required factory maintenance. Since May 11, 2018 the Cs clock has served as the time base for APOLLO.

\begin{table}
  \caption{\label{tab:clock}APOLLO timebase configuration changes}
  \begin{tabular}{ll}
    \hline\hline
    Date & Notes \\
    \hline\hline 
2016-02-12 & Start of Universal Counter clock comparison between Cs and XL-DC \\
2017-01-04 & Cs switched in as timebase for APOLLO ranging \\
2017-02-03 & Back to XL-DC (Cs clock issue) \\
2017-02-15 & Cs resumes as timebase \\
2017-12-21 & Back to XL-DC (Cs clock sent to manufacturer for repair)\\
2018-05-11 & Cs back in service as timebase \\
    \hline
    \end{tabular}
\end{table}

\begin{table}
  \caption{\label{tab:hardware}APOLLO hardware events}
  \begin{tabular}{ll}
    \hline\hline
    Date & Notes \\
    \hline\hline
2006-04-07 & Begin science operation using 30\,$\mu$m APD elements \\
2010-11-07 & Install 40\,$\mu$m APD (system suffered electrical trauma during exchange) \\
2012-04-07 & Change APD gate width; timing uncertainty improved\\
2013-09-07 & New APD bias/readout electronics; same 40\,$\mu$m APD \\
2016-09-12 & First ranges with ACS data \\
2019-08-21 & Replace APD with best 30\,$\mu$m device from new batch \\
2019-12-18 & Adjust optics to better center pupil image on APD; \\
           & no expected impact on timing \\
    \hline
    \end{tabular}
\end{table}

Table~\ref{tab:hardware} lists other significant changes to the APOLLO system, including those that could impact system timing and range measurement accuracy. Through a collaborative arrangement, we have been able to borrow and characterize $4\times 4$ APD arrays from MIT Lincoln Laboratory in 20, 30, and 40\,$\mu$m element sizes.  We began science operations using a $30\,\mu$m device from the original batch. In November 2010, we upgraded to an anti-reflection-coated $40\,\mu$m device from a new batch with lower dark current, easing tight alignment requirements. During this upgrade, the system experienced an electrical trauma and thus several system components were replaced. The resulting system timing was degraded relative to that before November 2010. In April 2012, we discovered that activating the APD detector 20\,ns earlier significantly improved the system timing. We attribute the improved timing to the settling of a transient prior to the arrival of lunar return photons. In September 2013, we upgraded the APD electronics, replacing the transistor-based scheme and coax transmission line couplings to daughter cards \cite{murphyAPOLLO2008} with a monolithic board solution using the Texas Instruments THS3001CD 420-MHz high-speed current-feedback amplifier.  The result was a faster ($<2$\,ns) rise time on the avalanche signal and better timing performance, contributing $< 30$\,ps to jitter. On 2016 September 12, we took the first lunar range measurements with simultaneous ACS calibration. In August 2019, we installed a 30\,$\mu$m device from the new batch that had an even lower dark rate and better timing uncertainty than the 40\,$\mu$m device. See Ref.~\cite{janaThesis} for studies of timing uncertainties in APD devices.

Each normal point in the APOLLO archive contains a reported uncertainty $\sigma_i$ that is based only on data from that 3--10 minute time span. By analyzing round-trip-time (RTT) data over longer periods of time, we found that the scatter exceeded expectations based on the $\sigma_i$ values alone. In particular, we compared RTTs from a single APD channel at a time to the weighted mean RTT from all channels and looked for evidence of systematic timing offsets. Details of this analysis process are provided in Ref.~\cite{apolloReduction}. We account for this excess noise by adding, in a root-sum-squared sense, an empirically determined systematic uncertainty $\sigma_{sys}$ to obtain a more accurate uncertainty estimate $\sigma_{i,tot}$ for each normal point, where $\sigma_{i,tot} \equiv \sqrt{\sigma_{i}^2 + \sigma_{sys}^2}$.
A single value of $\sigma_{sys}$ is assigned for each hardware period (see Table~\ref{tab:rss}).

\begin{table}
  \centering
  \caption{\label{tab:rss}Recommended systematic timing uncertainties for each APOLLO hardware period. Note: there are three exceptions to the table below. The normal point with timestamp 20120407T074405 should use $\sigma_{sys}=60$\,ps, and the normal points with timestamps 20120406T062935 and 20120406T063650 should use $\sigma_{sys}=23$\,ps.}
  \begin{tabularx}{0.8\textwidth}{XXrr}
    \hline\hline
    Start Date & End Date & \multicolumn{2}{c}{$\sigma_{sys}$} \\
               &          &      [ps]         & [mm one-way] \\ 
    \hline\hline
  2006-04-07 & 2010-10-30 & 17  & 2.5 \\ 
  2010-11-19 & 2012-04-06 & 60  & 9.0 \\ 
  2012-04-07 & 2013-09-01 & 23  & 3.5 \\  
  2013-09-30 & 2016-08-25 &  0  &   0 \\  
  2016-09-12 & 2020-12-27 &  0  &   0 \\  
  \hline
    \end{tabularx}
\end{table}

\begin{figure}
  \centering
  \includegraphics[width=\textwidth]{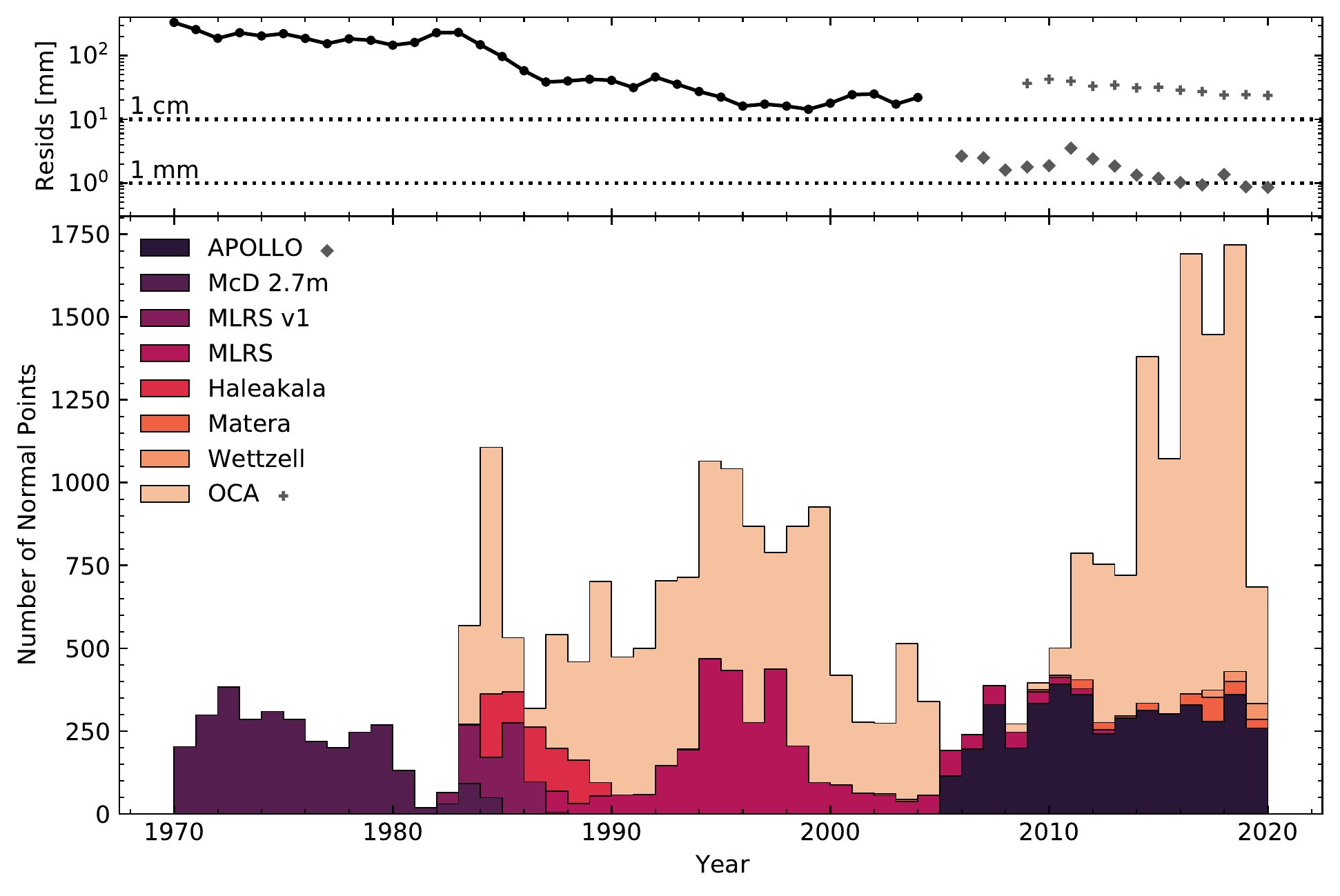}
  \caption{\label{fig:npPerYear}\textit{Bottom}: Stacked histogram showing the number of NPs produced per year by each lunar ranging station. The main contributors to the LLR dataset are McDonald/MLRS \cite{1996IAUS..172..409S}, OCA \cite{OCAIR}, and APOLLO (this work). \textit{Top:} Data-Model agreement (line with points) as quantified by the RMS of the post-fit residuals of LLR data with the JPL solar system ephemeris model (data adapted from \cite{2009IJMPD..18.1129W}). The post-fit residuals improved gradually during the 1970s due to the McDonald data, and more rapidly in the mid-1980s thanks to the start of (higher accuracy) ranging at OCA and the transition from McDonald to MLRS. Model and data inaccuracies both contribute to these residuals. To date, no ephemeris model is accurate at the millimeter level, so from 2006 onward, we show instead the annual median nightly normal point uncertainty for APOLLO ($\blackdiamond$) and OCA ($\usym{271A}$). For APOLLO, this quantity is an order of magnitude better, and better than 1\,mm for four of the last five years.}
\end{figure}

\section{The APOLLO LLR data set}\label{sec:data}
\subsection{General overview}
The global LLR data archive spans more than 50 years, beginning with range measurements made just two weeks after the Apollo 11 astronauts deployed the first retroreflectors on the lunar surface. Because the range link is weak (inversely proportional to the fourth power of the earth-moon separation, with overall attenuation factors on the order of $10^{-17}$), very few ranging stations on Earth have LLR capability. Figure~\ref{fig:npPerYear} shows the number of NP generated per year at each LLR station since 1970. Prior to APOLLO, two groups produced the bulk of the LLR dataset: the McDonald Laser Ranging Station (MLRS) in Texas, USA\footnote{LLR in Texas began on the McDonald 2.7\,m telescope (McD 2.7m), and then transitioned to a dedicated facility called the McDonald Laser Ranging Station (MLRS v1) which was subsequently relocated (MLRS).} \cite{1996IAUS..172..409S}, and the Observatoire de la C\^{o}te d'Azur (OCA) in Grasse, France \cite{OCAIR}. The figure also shows the onset of APOLLO operation in 2006. From April 2006 through the end of 2020, APOLLO produced 4296 NPs (35\% of all normal points in that time period). APOLLO data are freely available online \cite{np_dataset}, and are archived by the International Laser Ranging Service \cite{ilrs}. Figure~\ref{fig:npPerYear} also shows the median nightly uncertainty for each year of APOLLO and OCA operation, with $\sigma_{sys}$ included for the APOLLO data. The nightly uncertainty is defined in Section~\ref{sec:accuracy}. APOLLO's median nightly normal point uncertainty is below 4\,mm for all 15 years, below 2\,mm for 11 of the 15 years, and sub-mm for 4 of the last 5 years.

Because of its superior ranging accuracy, APOLLO's impact on fundamental physics is even more substantial than suggested by its fractional share of NPs. As a specific example, five years of APOLLO data alone produces science results (e.g. constraints on the equivalence principle) comparable to those based on the full archive of historical LLR data \cite{Muller:2012sea}.

\begin{figure}[t]
  \centering
  \includegraphics[width=0.31\textwidth]{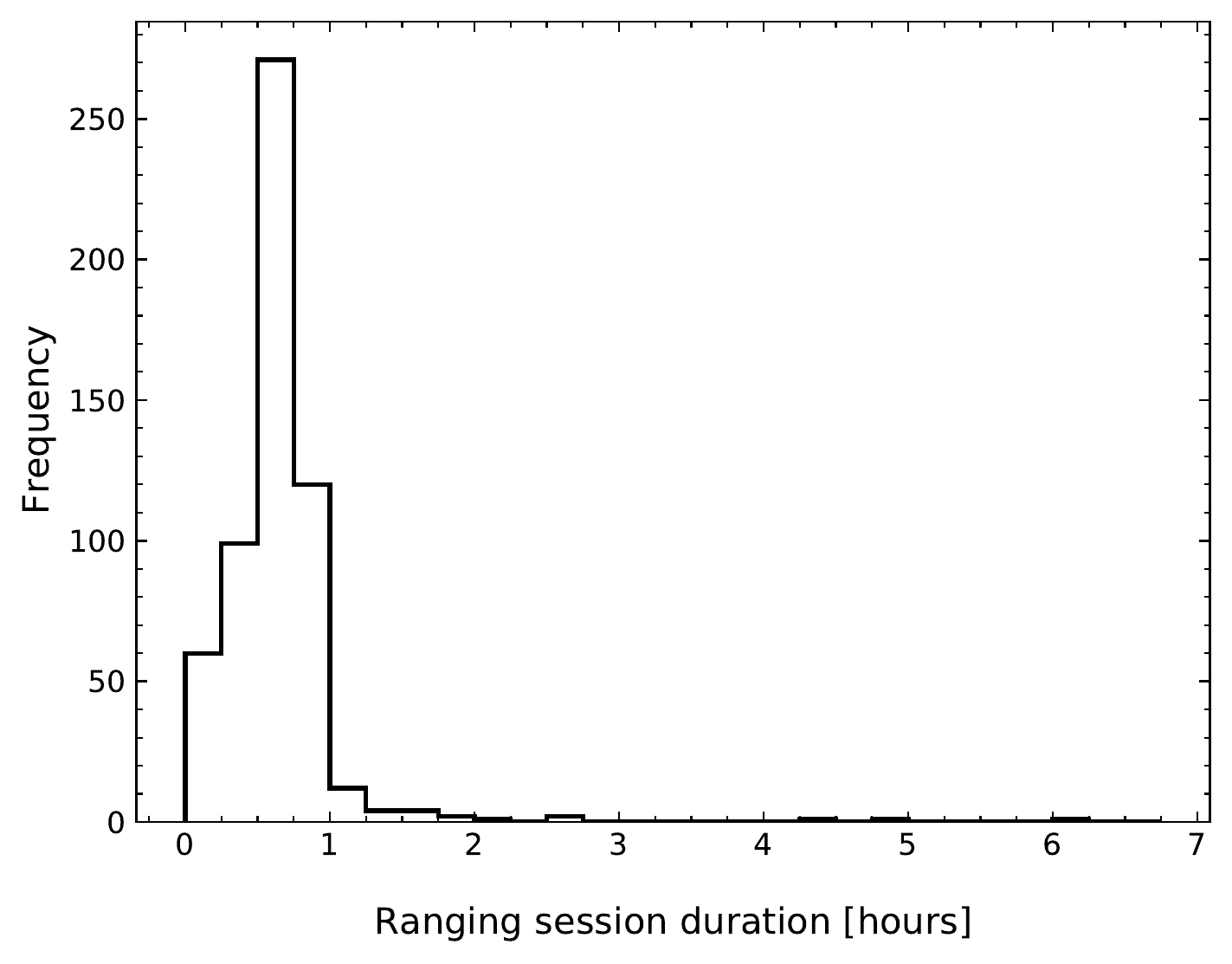}
  \includegraphics[width=0.31\textwidth]{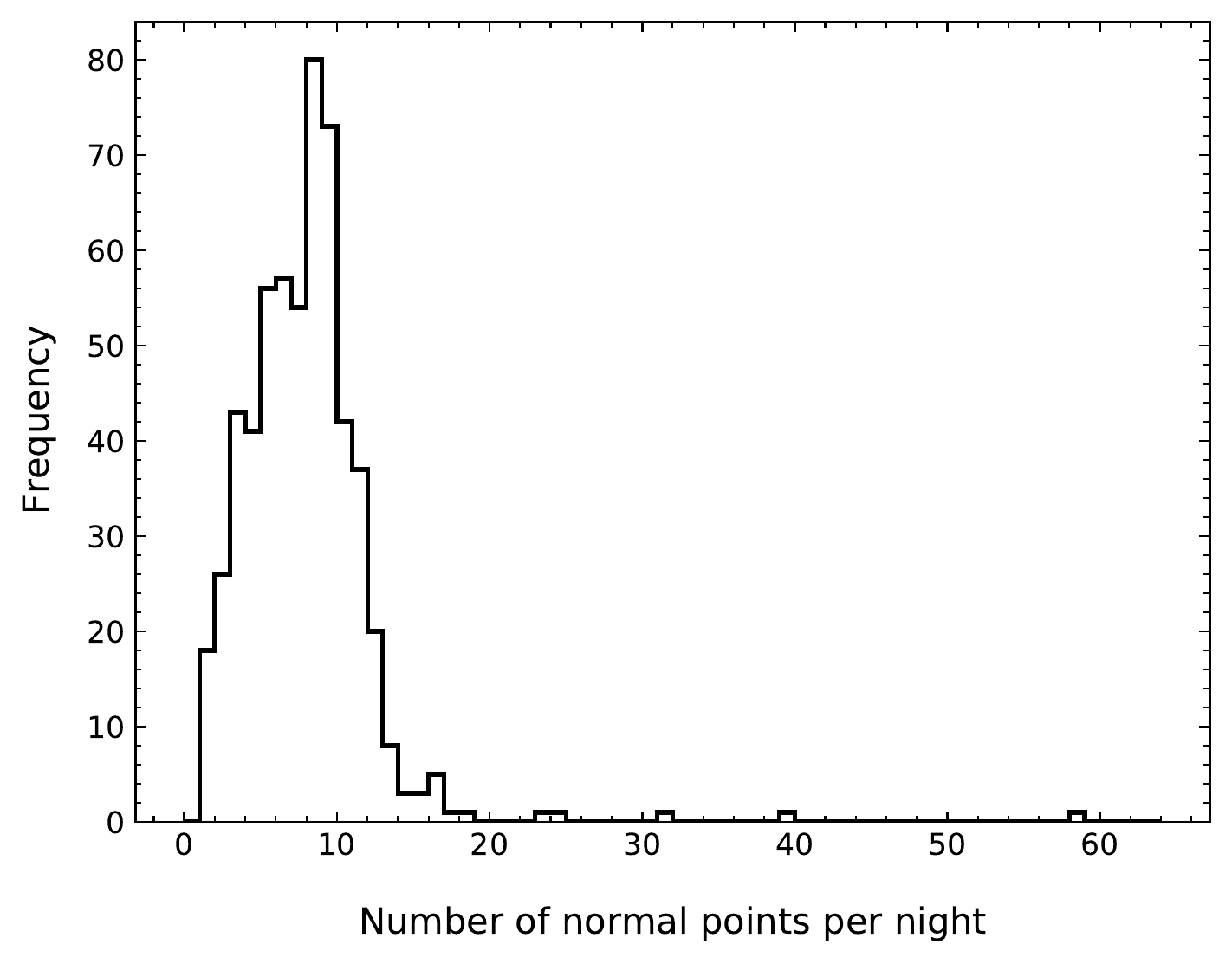}
  \includegraphics[width=0.31\textwidth]{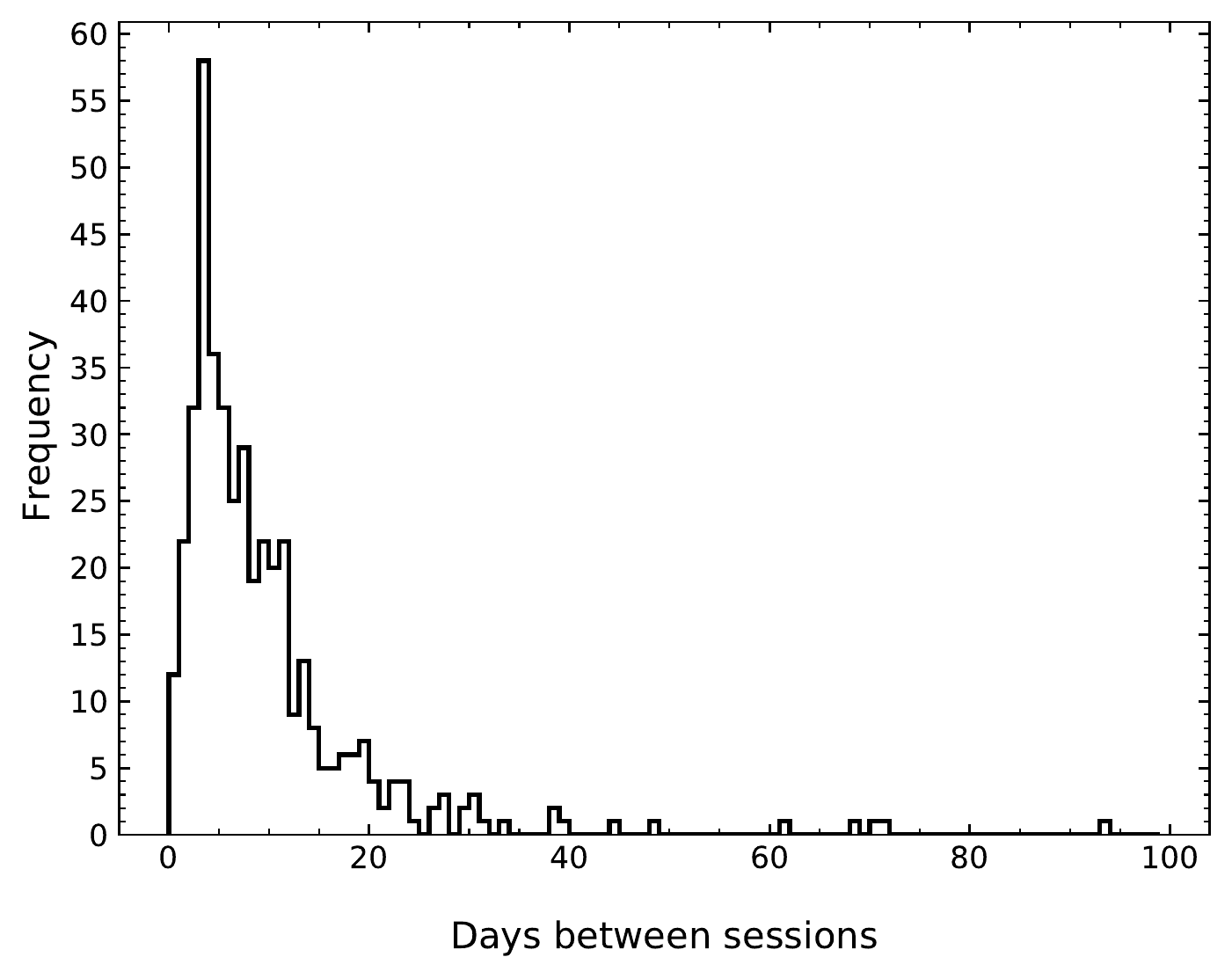}
  \caption{\label{fig:daysBetweenSessions} \textit{Left}: Distribution of ranging session duration. \textit{Middle:} Distribution of the number of NPs acquired per ranging session. A typical ranging session lasts under one hour and generates up to 10 NP. \textit{Right}: Distribution of time in days between successful APOLLO ranging sessions (in this context, ``successful'' means at least one NP was acquired). Typically, six ranging sessions are scheduled per lunation, and about one-third of the ranging sessions are lost to poor weather. The typical time between successful ranging sessions is 4--5 days. This provides good coverage over the period of the lunar orbit.}
\end{figure}

\begin{figure}[t]
  \centering
  \includegraphics[width=0.9\textwidth]{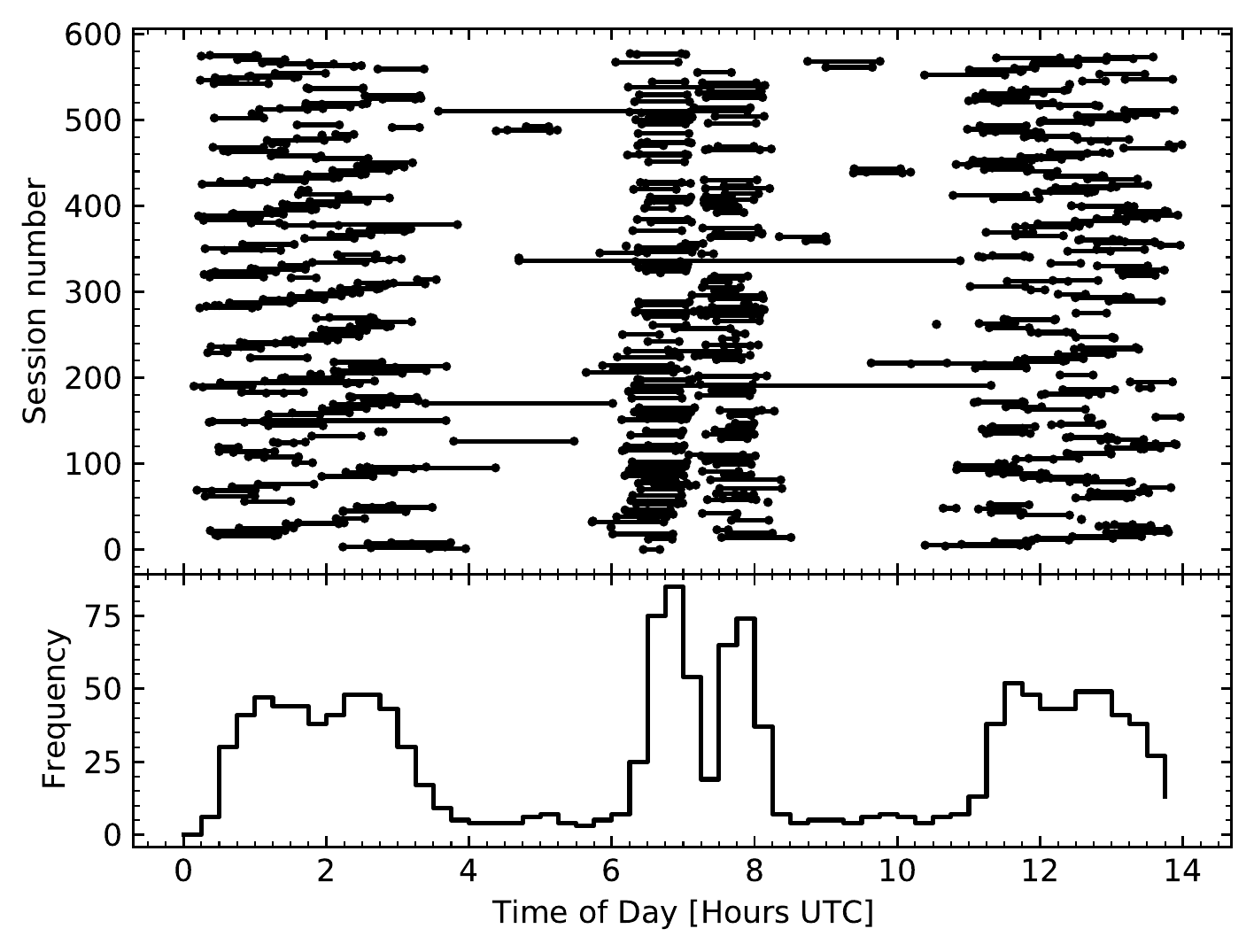}
  \caption{\label{fig:timeOfNight}\textit{Top:} Successful APOLLO ranging sessions, grouped by time of day in hours UTC. A horizontal line segment connects the time of the first and last NP from a given ranging session (``session number'' increments by one each ranging session, and is a non-linear measure of time). The evening start and morning end times show annual trends. \textit{Bottom:} Distribution of APOLLO ranging session times in hours UTC (bin width of 15 minutes) obtained by projecting the top plot onto the time axis. APO is located in the Mountain Time Zone of the US, which is UTC-7 (UTC-6 when observing daylight savings time).}
\end{figure}

Figure~\ref{fig:daysBetweenSessions} shows that a typical APOLLO ranging session lasts less than an hour and yields up to 10 NPs. On five occasions, APOLLO has undertaken extended ranging sessions (e.g. during lunar eclipses) that have produced more than 20 NPs per session. Because the APO 3.5m telescope is a shared-use facility, APOLLO cannot range every night. Instead, APOLLO purchased enough time on the telescope for six ranging sessions per lunation (the 29.53 day synodic period of the moon). Typically one-third of those allocated ranging sessions are lost to poor weather (telescope cannot open) or marginal weather (telescope can open and ranging is attempted, but the observing conditions are poor enough that no reflectors are successfully acquired). In addition, the APO 3.5m telescope undergoes a ``summer shutdown'' during New Mexico's monsoon season, and APO staff use that time for telescope engineering and maintenance. On occasion, APOLLO has forfeited a ranging session due to equipment malfunction. Figure~\ref{fig:daysBetweenSessions} shows that 4--5 days typically elapsed between successful ranging sessions. Telescope scheduling considerations also restrict the times of night that APOLLO can range. The APO 3.5m telescope is scheduled in half-night intervals, often with a change of instrument between sessions. APOLLO ranging sessions tend to take place at the start of the first session, at the end of the second, or between the two. Figure~\ref{fig:timeOfNight} clearly shows this clustering of observations near local evening, midnight, and morning.

\begin{figure}[tbh]
  \centering
  \includegraphics[width=0.9\textwidth]{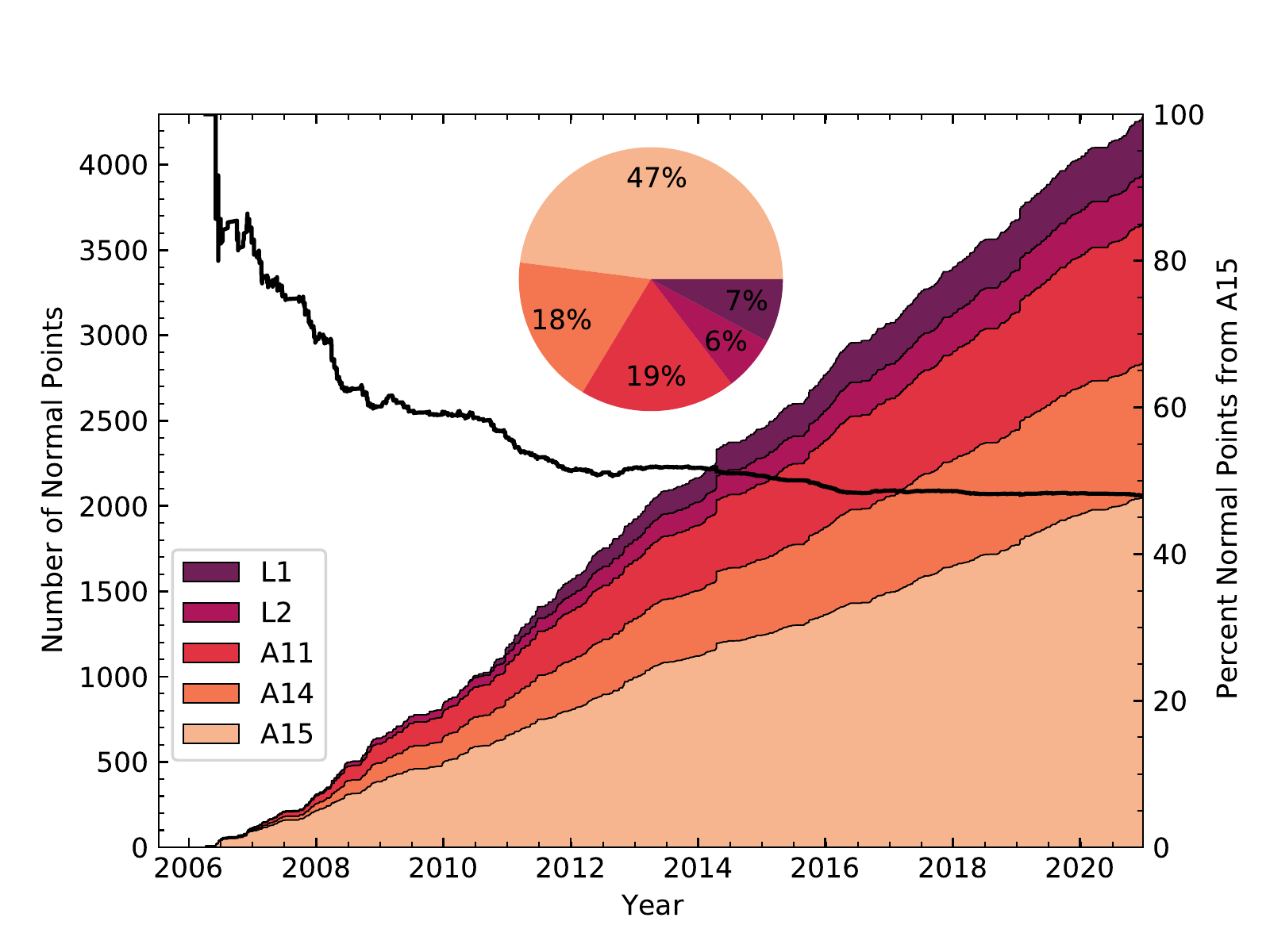}
  \caption{\label{fig:npPerTime}Number of APOLLO NPs per time (stacked plot, left ordinate), disaggregated by lunar reflector. The discovery of Lunokhod 1 by APOLLO is apparent in 2010, as are the annual summer shutdowns at APO (flat regions), and four extended ranging sessions (vertical steps) during lunar eclipses in December 2010, April 2014, September 2015, and January 2019. Also provided are the fraction of NPs on A15 (solid black curve, right ordinate) and a pie chart showing the distribution of lunar reflectors in the APOLLO data set. In the early years, APOLLO relied heavily on the stronger signal from A15, but that dependence has steadily decreased. Overall, A15 measurements make up less than half of the APOLLO dataset.}
\end{figure}

\subsection{Per-reflector statistics}
Figure~\ref{fig:npPerTime} shows the growth of the APOLLO dataset over time, disaggregated by lunar reflector. APOLLO's (re)discovery of the Lunokhod 1 reflector (L1) in 2010 \cite{Murphy:2010ua} is visible in that plot, as are the APO 3.5m telescope summer shutdowns (flat regions of the curves), and several eclipse observations during which $>20$ NP are produced in a single night. The inset pie chart shows that just under half of all APOLLO NPs are measured to Apollo 15 (A15), approximately 20\% each on Apollo 11 and 14 (A11 and A14), and less than 10\% each on Lunokhod 1 and 2 (L1 and L2). A15 is overrepresented in the data because it is the largest of the five arrays (three times larger than A11 and A14), which produces a correspondingly larger return signal. Thus ranging sessions at APOLLO (and other ranging stations) typically begin with a search for A15 before proceeding to other reflectors. APOLLO also typically closes out each ranging session by ranging to A15 to provide a good lever arm for determining Earth rotation. Over time, however, APOLLO has reduced its reliance on the Apollo 15 array (Figure~\ref{fig:npPerTime}, black curve) to the point where the majority of APOLLO NPs come from other reflectors.

\begin{figure}
  \centering
  \includegraphics[width=0.9\textwidth]{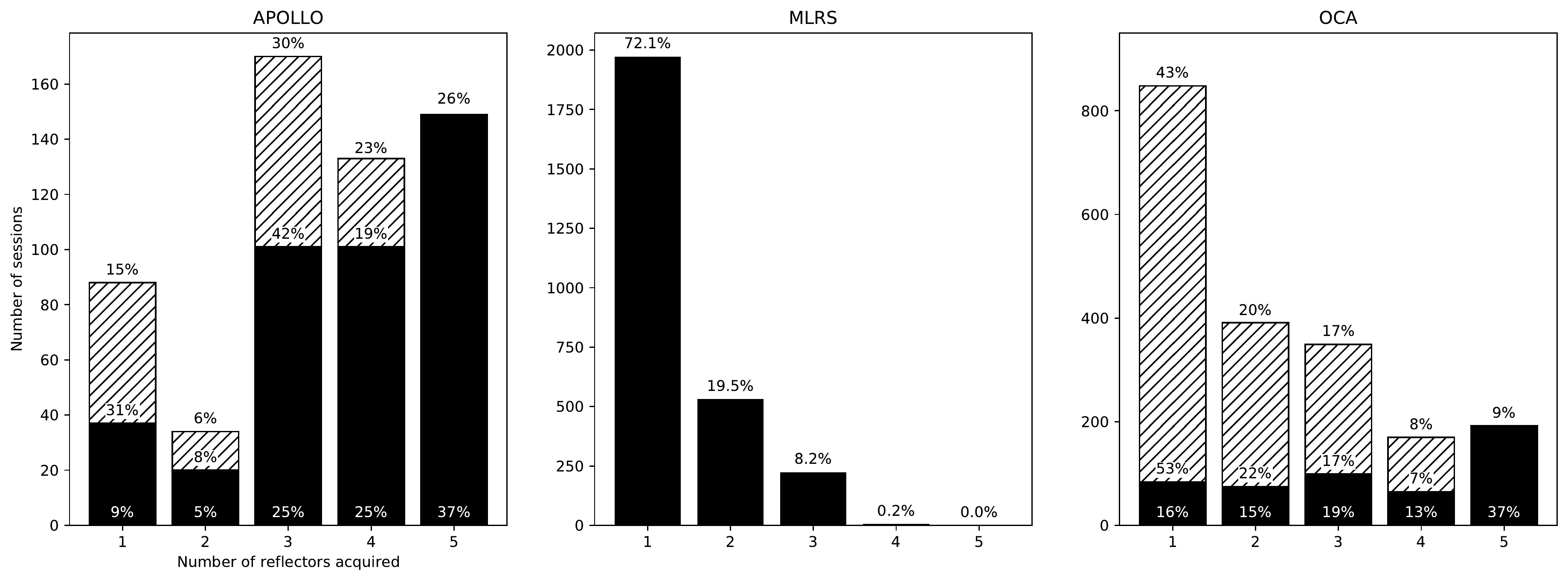}
  \caption{\label{fig:reflPerSession}Number of lunar reflector arrays acquired in a single ranging session for APOLLO (tens of minutes to an hour), MLRS, and OCA. \textit{Left}:  Prior to the rediscovery of L1 in 2010 (hatched), APOLLO most often acquired three arrays. Since the rediscovery of L1 (solid black), the most likely outcome is that APOLLO acquires all five reflectors during a single ranging session, and 87\% of the time acquires three or more. The percentages listed at the bottom of the black bars, bottom of the hatched bars, and atop the bars are for post-L1 data, pre-L1 data, and the full APOLLO dataset, respectively. Due to rounding errors, the percentages may not sum to 100\%. \textit{Middle}: MLRS typically ranged to just one reflector per night and only acquired three or more reflectors 8.4\% of the time. \textit{Right}: OCA data before implementing infrared ranging (hatched) and after (black). Since the upgrade, OCA acquired all five arrays in a single night 37\% of the time and three or more 69\% of the time. Percentage labels at the bottom of the black bars, in the hatched bars, and atop the bars are for post-IR data, pre-IR data, the full OCA dataset, respectively.}
\end{figure}

APOLLO achieves millimeter precision largely due to a high photon return rate, which also enables the consistent acquisition of signal from multiple reflector arrays in a short period of time (tens of minutes to an hour). Figure~\ref{fig:reflPerSession} shows the distribution of the number of distinct retroreflector arrays from which APOLLO acquires signal in a single ranging session. Rarely does APOLLO acquire only two reflectors (if the ranging conditions are good enough to acquire two, then they are good enough to acquire three or more). In fact, since the rediscovery of L1, APOLLO acquires four or five reflectors 62\% of the time, and three or more (required to constrain lunar orientation and tidal distortion) 87\% of the time.

APOLLO clearly outperforms MLRS in this regard: 72\% of the time MLRS only acquired one reflector, and only 8.4\% of the time was it able to acquire three or more. Although on most nights OCA has only acquired one reflector, a 2015 instrument upgrade to enable infrared ranging (1064\,nm) significantly improved their return rate and ability to acquire multiple reflectors in quick succession \cite{OCAIR}. Figure~\ref{fig:reflPerSession} shows that since that upgrade, OCA acquired 3 or more reflectors 69\% of the time, and is most likely to acquire all five reflectors on a given night.

\begin{figure}
  \centering
  \includegraphics[width=0.4\textwidth]{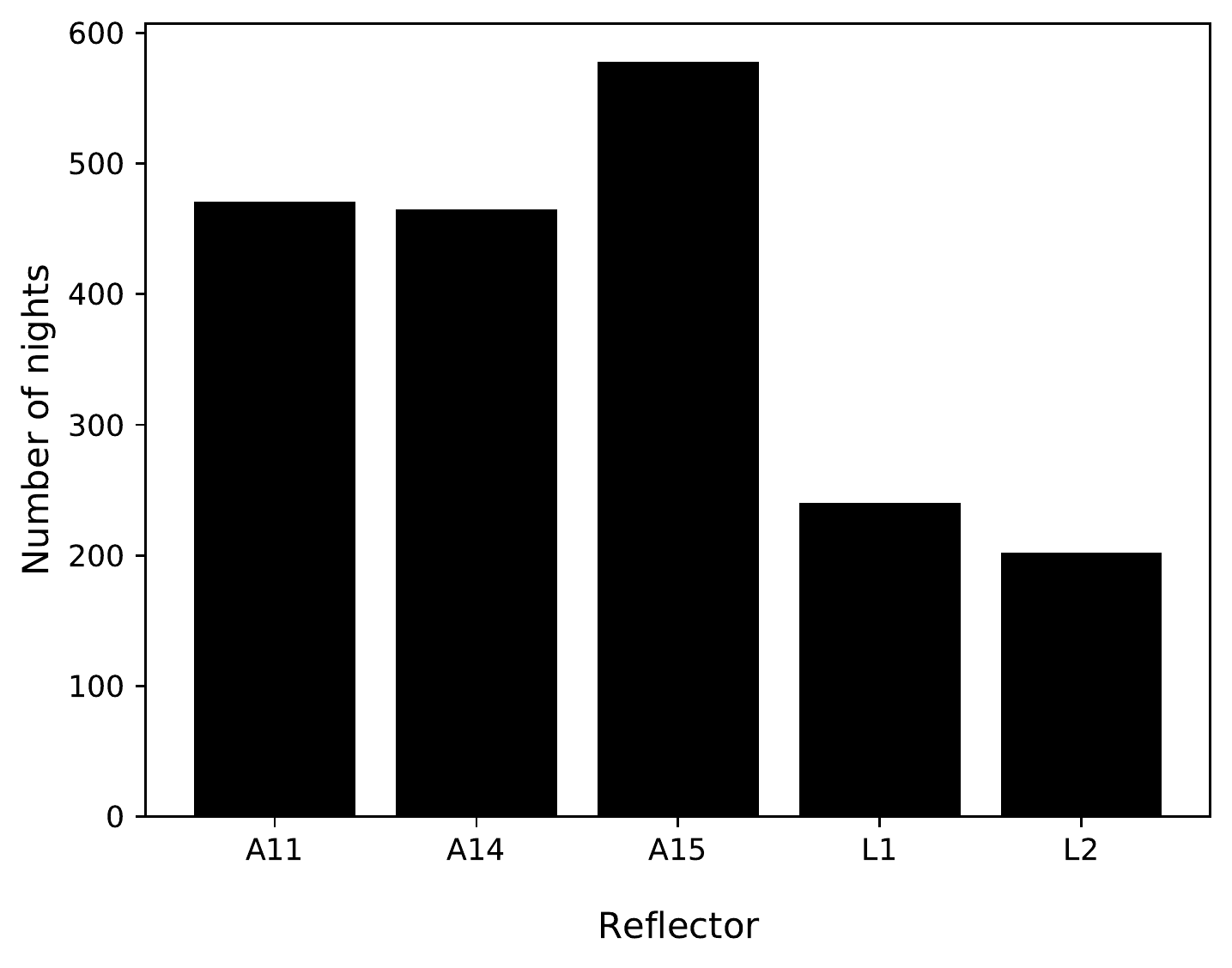}
  \caption{\label{fig:nightsPerRefl}Number of successful APOLLO observing sessions (``nights'') for each reflector.}
\end{figure}

In Fig.~\ref{fig:nightsPerRefl}, we indicate how often (number of ranging sessions) APOLLO successfully acquired a given array. It was rare for A14 to be acquired without A11 (consistent with the strategy of starting with A15 and then moving onto A11 and A14, and with the fact that APOLLO rarely acquires only two reflectors). Both A11 and A14 were acquired more than twice as often than the Lunokhod arrays, in large part because L1 and L2 are much more sensitive to solar illumination -- the metallized backing of the retroreflectors absorbs sunlight, thereby creating thermal distortions that significantly degrade the photon throughput.

\subsection{Ranging accuracy}\label{sec:accuracy}
We use several different metrics to quantify the accuracy of the APOLLO range measurements. The main conclusion in all cases is that the accuracy of APOLLO normal points is at the millimeter scale.

\begin{figure}
  \centering
  \includegraphics[width=0.9\textwidth]{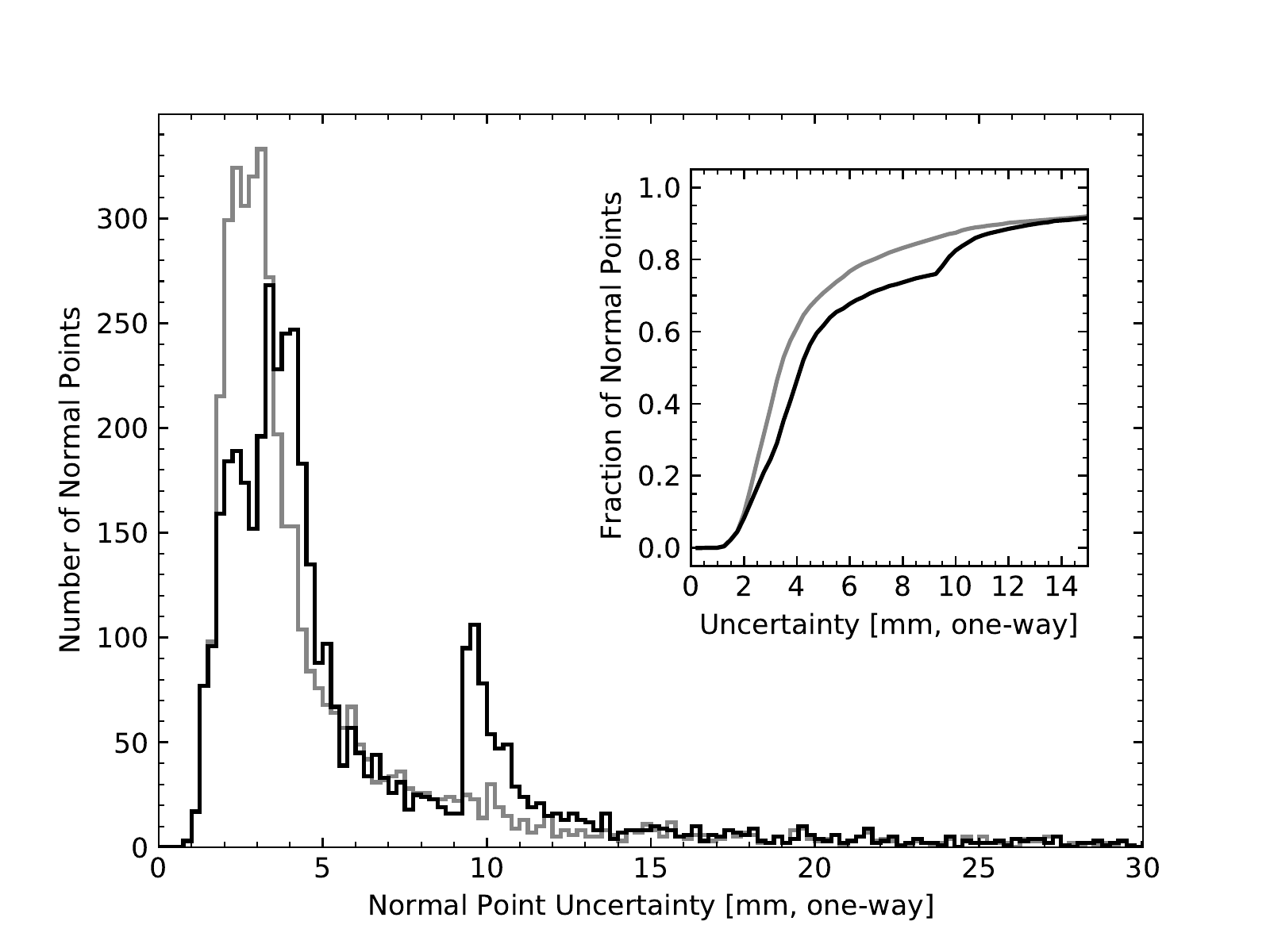}
  \caption{\label{fig:npUncertAPOLLO} Distribution of APOLLO normal point uncertainties $\sigma_i$ (gray) showing a typical uncertainty of a few millimeters. The black line shows the distribution of $\sigma_{i,tot}$ which accounts for systematic timing uncertainties, as described in Sec.~\ref{sec:instrument}. (Inset): Cumulative distribution of these distributions, showing that half of the normal points have an uncertainty below 3.4\,mm (for $\sigma_i$) and 4.2\,mm (for $\sigma_{i,tot}$).}
\end{figure}

In Fig.~\ref{fig:npUncertAPOLLO}, we show the distribution of the uncertainties of each NP in the full APOLLO dataset. We plot both the raw uncertainty $\sigma_i$ for each NP as well as $\sigma_{i,tot}$, which accounts for empirically determined systematic timing errors. The median raw uncertainty is 3.4\,mm, while the distribution of $\sigma_{i,tot}$ has a median of 4.2\,mm.

\begin{figure}
  \centering
  \includegraphics[width=\textwidth]{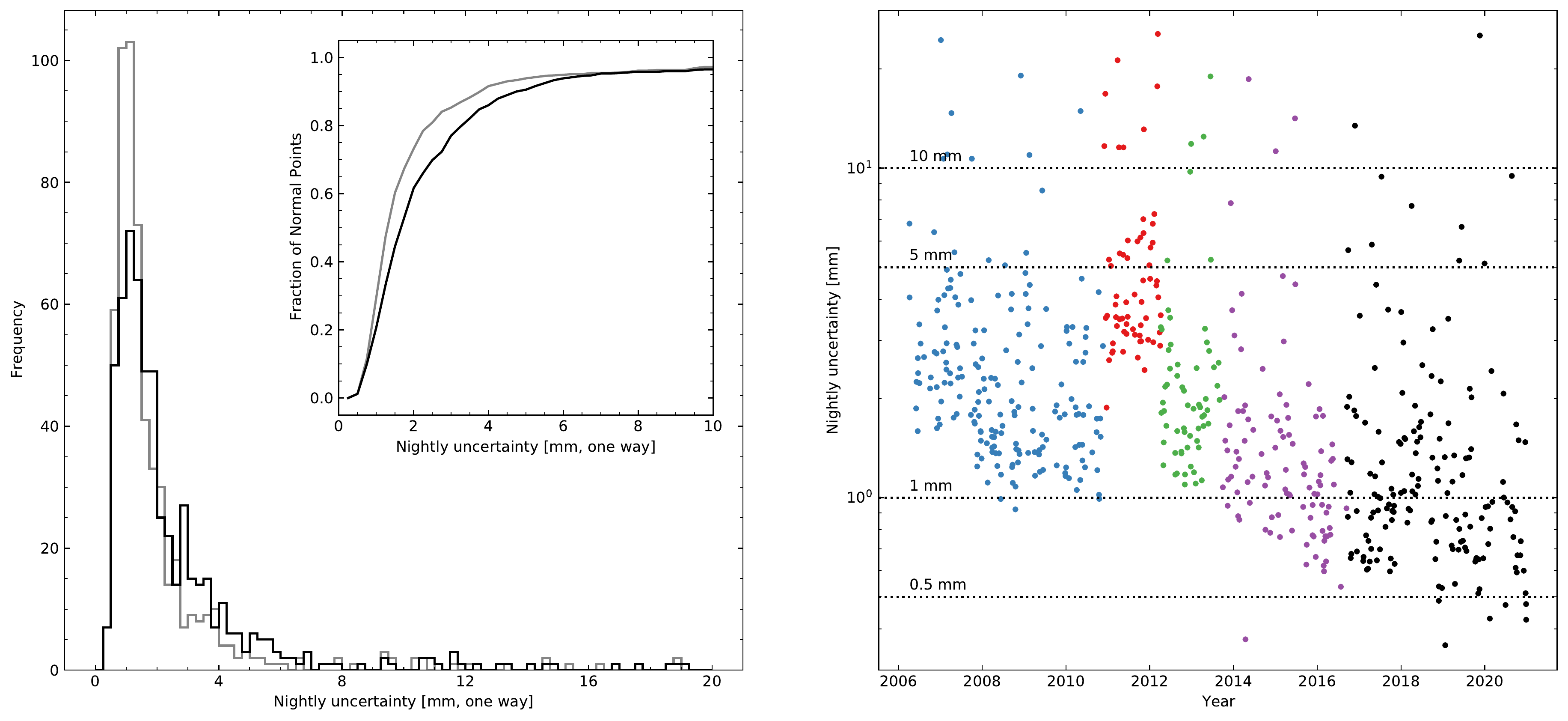}
  \caption{\label{fig:nightlyUncert} \textit{Left}: Distribution of nightly uncertainties $\sigma_N$ (gray) and $\sigma_{N,tot}$ (black). Typical nights contain $\sim$10 individual NPs (see Fig.~\ref{fig:daysBetweenSessions}, middle). The inset shows the associated cumulative distributions. The median nightly uncertainties are 1.3\,mm and 1.7\,mm for $\sigma_N$ and $\sigma_{N,tot}$, respectively. \textit{Right}: Nightly uncertainty $\sigma_{N,tot}$ as a function of time for the duration of the APOLLO experiment colored by the five different hardware periods listed in Table~\ref{tab:rss}. The degraded accuracy from December 2010 through April 2012 is apparent (red points). The median of $\sigma_{N,tot}$ for the most recent period (black points) is 1.0\,mm.}\end{figure}

Given that most potential physics signals of interest in the LLR data involve timescales that are much longer than an hour, it is also useful to report the ``nightly uncertainty'' which takes into account all of the NPs acquired in a single ranging session, typically lasting 45 minutes (see Fig.~\ref{fig:daysBetweenSessions}). We define the nightly uncertainties $\sigma_N$ and $\sigma_{N,tot}$ in terms of the $k$ individual NP uncertainties from a single ranging session as:
\begin{equation}
\sigma_N       \equiv \sqrt{\frac{1}{\sum_{i=1}^k \sigma_i^{-2}  }} \qquad\qquad
\sigma_{N,tot} \equiv \sqrt{\frac{1}{\sum_{i=1}^k \sigma_{i,tot}^{-2}  }}
\end{equation}
Figure~\ref{fig:nightlyUncert} shows the distributions of $\sigma_N$ and $\sigma_{N,tot}$. The median nightly uncertainty is 1.3\,mm and 1.7\,mm for $\sigma_N$ and $\sigma_{N,tot}$, respectively. As described in Sec.~\ref{sec:instrument} and Table~\ref{tab:hardware}, APOLLO has undergone several hardware changes and upgrades over the past 15 years, with implications on ranging accuracy. Figure~\ref{fig:nightlyUncert} also shows the evolution of the nightly uncertainty $\sigma_{N,tot}$ over the duration of the APOLLO experiment. Since September 2016, the median total nightly uncertainty $\sigma_{N,tot}$ is 1.0\,mm.

\subsection{Lunar phase angle distribution}\label{sec:lunarphase}
\begin{figure}
  \centering
  \includegraphics[width=\textwidth]{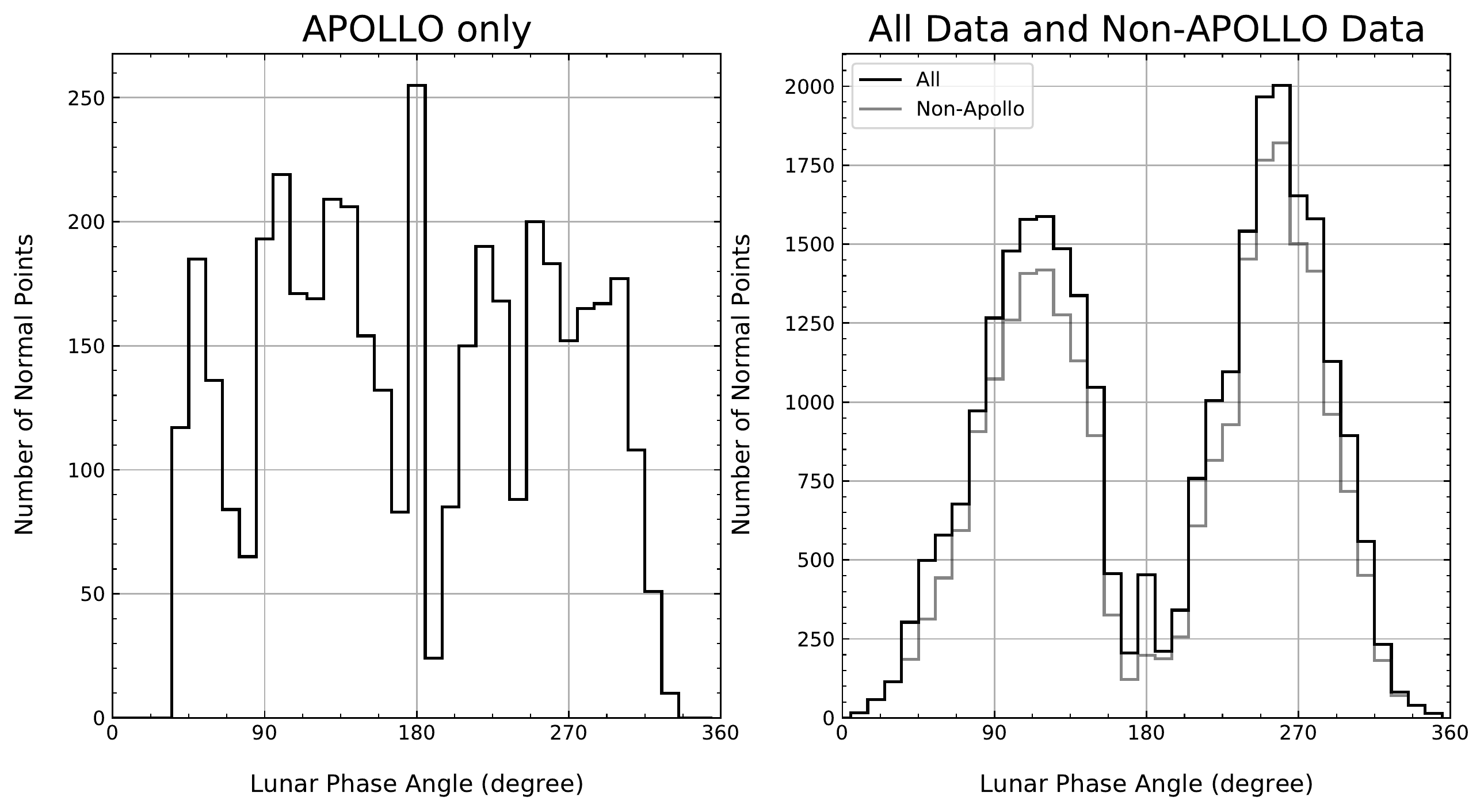}
  \caption{\label{fig:npLunarPhase} Distribution of NPs as a function of lunar phase angle. New moon is $0^\circ$ and $360^\circ$, and full moon corresponds to $180^\circ$. \textit{Left:} APOLLO data. The absence of measurements at new moon is due to telescope sun avoidance restrictions. A deficit of measurements near full moon is caused by the absorption of solar radiation by the lunar corner cubes, which degrades the range link throughput by an order of magnitude. The excess of measurements at full moon is due to extended ranging sessions during four lunar eclipses. \textit{Right:} Same plot but for data from all LLR stations (black) and non-APOLLO stations (gray) showing the systematic deficit of full and new moon observations relative to first and last quarter measurements (phase angles of $90^\circ$ and $270^\circ$).}
\end{figure}

As described previously, LLR is sensitive to an SEP violation through the synodic perturbation to the lunar orbit that would be produced by a differential acceleration of the earth and moon toward the sun. This periodic signature peaks at new and full moon, but the LLR data archive is sparsely populated at those phases. Figure~\ref{fig:npLunarPhase} shows the distributions of LLR observations with respect to the mean elongation $D$ of the moon from the sun. We clearly see the absence of observations at new moon ($D=0^\circ$) due to telescope Sun avoidance criteria, as well as a deficit of successful range measurements near full moon ($D=180^\circ$). OCA and APOLLO are the primary contributors to full-moon observations, with most of APOLLO's full-moon NPs obtained on 4 nights of extended lunar eclipse observations. 

\section{Conclusion}\label{sec:conclusion}
Since April 2006, APOLLO has been measuring the earth-moon distance, achieving millimeter-scale accuracy as a probe of fundamental physics as well as lunar properties. High return rates and a multi-pixel sensor allow APOLLO to routinely range to three or more lunar arrays during a single hour-long ranging session. Successful ranging is routinely achieved during full moon, though thermal distortions of the lunar retroreflectors degrade the signal throughput. Incorporation of the ACS into the APOLLO system in September 2016 allows for in-situ system timing calibration and provides reliable measurements of NP uncertainties. Future work on data reduction could lead to improved NPs from the same set of raw range measurements. In particular, the current analysis rejects all range measurements for which more than one APD pixel received lunar returns from a single laser shot. Relaxing this restriction, while taking care to avoid range timing bias~\cite{ericThesis}, could significantly improve the APOLLO NP archive. In particular, including laser shots in which more than one lunar photon was detected would increase the number of shots used in the analysis by 30\%, and the number of registered lunar photons by 79\%. The operation of APOLLO was transferred to NASA Goddard Space Flight Center in January 2021. This data release is the final one from the team that built the instrument.

\section*{Acknowledgments}
John Chandler provided data quality checks via data-model comparisons. Liliane Biskupek and Vishnu Viswanathan shared and helped to interpret data from other LLR observatories, including OCA. This work is based on access to and observations with the Apache Point Observatory 3.5-meter telescope, which is owned and operated by the Astrophysical Research Consortium. The avalanche photodiode detectors were provided by Lincoln Laboratory. This work was jointly funded by the National Science Foundation (PHY-0602507, PHY-1068879, PHY-1404491 and PHY-1708215) and the National Aeronautics and Space Administration (NNG04GD48G, NAG81756, NNX12AE96G, NNX15AC51G and 80NSSC18K0482). JBRB acknowledges the support of the NASA Massachusetts Space Grant (NNX16AH49H).

\section*{References}
\bibliographystyle{iopart-num}
\bibliography{apollo}

\end{document}